\documentstyle[11pt]{article}
\begin{document}

\begin{flushright}
Liverpool Preprint: LTH 388\\
 hep-lat/9611011\\
 8 November 1996\\
 \end{flushright}
  
\vspace{5mm}
\begin{center}
{\LARGE\bf Hybrid mesons from quenched QCD   }\\[10mm] 
{\large\it UKQCD Collaboration}\\[3mm]
 
{\bf  P. Lacock, C. Michael}\\

{Theoretical Physics Division, Department of Mathematical Sciences, 
University of Liverpool, Liverpool, L69 3BX, U.K.}\\[2mm]

{\bf P. Boyle, P. Rowland}\\

{Department of Physics \& Astronomy, University of Edinburgh, 
Edinburgh EH9 3JZ, Scotland}\\

\end{center}

\begin{abstract}

We use lattice methods to evaluate from first principles the spectrum of
hybrid mesons produced by gluonic  excitations  in quenched QCD
with quark masses near the  strange quark mass.   For  the  spin-exotic 
mesons with $J^{PC}=1^{-+},\ 0^{+-}$, and $2^{+-}$ which are not present
in the quark model, we determine the lightest state to be $1^{-+}$ with 
mass of 2.0(2) GeV.

\end{abstract}

One of the goals of quantitative studies of QCD is to determine the 
masses and properties of states which are not allowed in the simple
quark model because they contain  gluonic excitations.  The prototype
for such a state is the glueball and accurate lattice
studies~\cite{ukqcd,gf11} have been able to pinpoint the  relevant mass
range for experimental study of the scalar glueball as around 1.6 GeV.
Another important area is the study of hybrid mesons which are $q
\bar{q}$ mesons with gluonic excitation.  Because the gluonic excitation
can introduce angular momentum, the most clear cut  signal for a hybrid
meson is to search for $J^{PC}$ quantum numbers not allowed  in the
quark model. These include $J^{PC}=1^{-+},\ 0^{+-}$, and $2^{+-}$. Again
lattice QCD is capable, from first  principles, of  establishing the
masses of these states.  

In full QCD, the quark pair creation and anihilation processes will lead
to  the mixing of glueballs with $q\bar{q}$ mesons and it is thus
appropriate to establish the glueball spectrum with these processes
turned off - the quenched approximation. This then acts as a guide for
mixing studies.  For  hybrid mesons, however, the quenched approximation
already allows mixing  between  $q\bar{q}$ mesons and hybrid mesons if
their $J^{PC}$ quantum numbers are non-exotic. This will be quite
difficult to untangle in lattice  studies, so we focus instead on the
exotic $J^{PC}$ cases.  Any mixing for these exotic $J^{PC}$ states
would have to be with  $q\bar{q}q\bar{q}$ mesons and such mixing is
turned off in the quenched approximation.  Thus, in the quenched
approximation,  we are able to determine unambiguously the exotic hybrid
energy levels given by the full  non-perturbative gluonic interaction.

The spectrum of hybrid mesons has been explored in lattice QCD for  many
years~\cite{liv,cmper} in the context of heavy quarks. In the quenched
approximation, the lightest hybrid level is found to be degenerate and
to contain  spin-exotic  mesons with $J^{PC}=1^{-+},\ 0^{+-}$, and
$2^{+-}$.  Since the quenched approximation does not reproduce exactly 
the experimental $c \bar{c}$ and $ b \bar{b}$ spectra,  the prescription
of  increasing the short distance component in the quenched
approximation potential yields the lattice prediction~\cite{cmper}  for
the lightest hybrid meson  excitation to be at 4.19(15) GeV for $c
\bar{c}$ and 10.81(25) GeV  for $b \bar{b}$.   These energy values lie
above the open $D \bar{D}$ and $B \bar{B}$ thresholds. An alternative 
procedure, within the quenched approximation, is to focus~\cite{sommerb}
on the energy  difference between the hybrid meson and the $B\bar{B}$
threshold. This suggests that the lowest hybrid level may lie below  the
threshold. If the hybrid states  were to lie  below these thresholds,
then they would be narrow resonances and so much  clearer to detect
experimentally.  These lattice results are for the quenched
approximation and do not include mixing  with 4-quark states. They also
they do not include spin-orbit effects which may be significant for
quarks of finite mass. This is one motivation for exploring hybrid 
levels on a lattice with realistic quark masses.

There have been many experimental claims of signals for  $J^{PC}$ exotic
mesons in light quark spectroscopy (by light we  mean composed of u, d,
s quarks). None of these claims are overwhelming~\cite{pdg,close}  and
guidance from QCD is sorely needed, as it was for glueball exploration.
With present computational resources, it is possible to explore this 
and here we present the results of the first systematic study. Details 
of our lattice techniques and our preliminary results from much lower
statistics  have been published earlier~\cite{prdhyb}. Another  lattice
collaboration is also engaged in a study of this area~\cite{ushyb}.

Spin-exotic hybrid mesons have gluonic components with non-trivial
angular  momentum. This can be modelled on a lattice by joining the
quark and antiquark  by a colour flux path (product of links) which is
not straight. Guidance in designing  suitable paths comes from  the
heavy quark studies where a U-shaped path (actually a difference of  the
form  $\, \sqcap - \sqcup$ ) between static quarks  was
found~\cite{liv,cmper} to yield the lightest hybrid excitation. We
construct hybrid meson operators for light quarks by joining the light
quarks by colour flux in that U-shaped orientation.  This involves an
extra quark propagator inversion, but we are able to  study a range of
different hybrid mesons using this extra inversion. Effectively the
technique is to employ a `white'  source on the lattice and to separate
the hybrid state of interest by choosing an appropriate operator for the
sink~\cite{prdhyb}.

\begin{table}
\begin{center}
\begin{tabular}{lclll}
 meson&  $J^{PC} $  & mass& $M/M_V$&$M_s/M(\phi)$\\
 multiplet &      &     $Ma$&  &\\ \hline
$\hat{\rho}$&$  1^{-+},\ 3^{-+} $ &0.95(7)&1.76(13) &1.95(13)\\
\^a$_0$&$ 0^{+-},\ 4^{+-}$  & 1.05(7)& 1.94(13)&2.16(13)\\
\^a$_2$&$  2^{+-},\ 4^{+-} $   & 1.26(13)& 2.33(24)&2.62(24)\\
\^a$_2$&$  2^{+-},\ 3^{+-} $   & 1.12(10)&2.07(18) &2.32(18)\\
\end{tabular}
\end{center}
 \caption{The masses of the hybrid mesons. The $J^{PC}$ values are the
lowest two values allowed by the lattice cubic symmetry.   The masses
are given in lattice units and as a ratio to the vector meson mass
$m_V$. The mass ratio  appropriate to   $s \bar{s}$ states is given in
the last column.}
 \end{table}

Details of the lattice evaluation and operator construction are given in
 ref.~\cite{prdhyb} and we summarise them here.  We used a $16^3 \times
48$ lattice at $\beta=6.0$ with the Wilson gauge  action for SU(3)
colour fields. This corresponds to a lattice spacing  of approximately
0.1 fm. The quark propagators were evaluated at hopping parameter 
$K=0.137$ using an SW-clover fermionic action with coefficient 
$c=1.4785$.  For our U-shaped source we took a separation of 6 lattice
units and we evaluated 350 local propagators from sources at $(0,0,0,0)$
and  at $(0,0,6,0)$.   For the sink at time $t$, we used U-shapes of
size 1, 3 and 6 lattice units. 

Using the same operator at source and sink (namely with size 6) gives an
 effective mass which is an upper bound on the ground state mass. In
order to estimate this ground state mass,  we evaluate the correlation
between our source operator and all three sink operators since this data
set allows  us to estimate the excited state contribution and so extract
the ground state  mass by making stable two state fits to these
correlations. The effective mass values for $\hat{\rho}$ are shown in the
Figure. Given that the  signal to noise ratio deteriorates rapidly with
increasing time separation $t$,  it is vital to use the small $t$ data
to extract the ground state mass. We  used correlated fits to as wide a
range of $t$ as possible consistent with  acceptable goodness of fit. A
correlated 2-state fit to $t$-range  2-10 was found to give the most
reliable determination of the  ground state mass. The statistical error
was evaluated using bootstrap. A systematic error from the uncertainty 
in this fitting procedure was determined by considering a range of other
fits  (by varying  the $t$-range, fixing the  excited state mass
excitation to 1.5 lattice units, using a 1-state fit only, using an
uncorrelated fit, fitting only subsets  of the data). In each case this
systematic error was found to be not larger  than the statistical  error
quoted.

\begin{figure}[p]
\vspace{14cm} 
\includegraphics{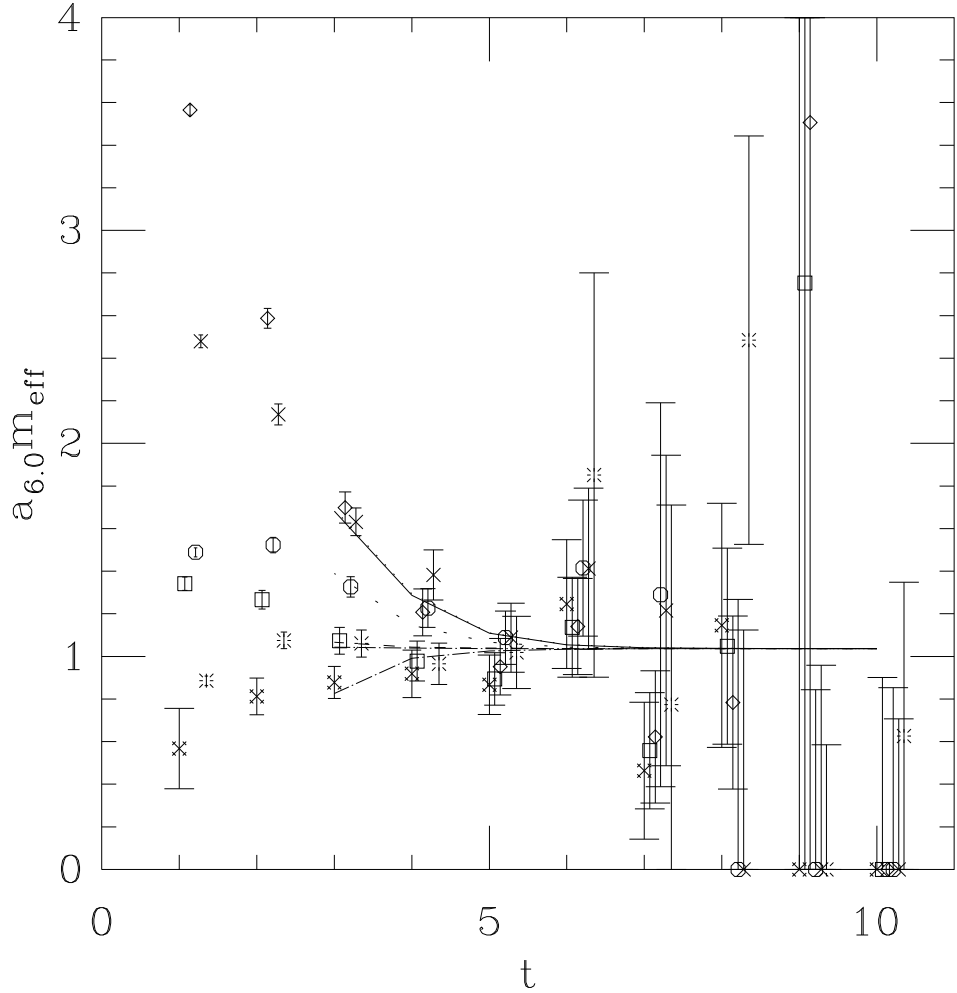}
 \caption{ The lattice effective mass for the $J^{PC}=1^{-+}$  hybrid
meson versus time separation $t$. The source used was a U-shaped path of
size  $6 \times 6$, while the sinks were combinations of U-shaped paths 
of size $6 \times 6$ ($\diamond$), $3 \times 3$ ($\Box$) and $1 \times
1$ (fancy cross). We also used closed colour loops at the source  of size
$6 \times 6$ ($\ast$), $3 \times 3$ (octagon) and $1 \times 1$
($\times$) with the above sinks. The simultaneous two state fit to 
all six  correlations is shown.
   }
\end{figure}

Our lattice results for the  $1^{-+},\ 0^{+-}$ and $2^{+-}$ exotic
hybrid mesons are given in the Table.  Note that correlation studied in
the last row  in the Table can have contributions from a non-exotic
$b_3$  meson as well as the exotic $2^{+-}$, so it is only included for
completeness.   Our results are consistent with the assumption that the
$1^{-+}$ state  is the lightest. Taking account of statistical
correlations, we find that the $0^{+-}$ state is heavier  than the
$1^{-+}$ state with a significance of 1 standard deviation. 

As a cross check we also studied these states using  a loop of colour
flux which starts and finishes at the same site to model the gluonic
excitation at the source. This closed gluon loop with non-trivial colour
has the advantage of creating hybrid states which acn be studied without
any additional quark  propagators beyond those usually used in hadron
spectroscopy. The disadvantage is that it  creates hybrid states with 
the quark and antiquark at the same location and this may be an
untypical component of the wavefunction. For the  $J^{PC}=1^{-+}$ state,
we used a  `magnetic'  colour loop ($L^{PC}=1^{+-}$)  and, as shown in 
the Figure, the effective  mass  from measuring 294 configurations is
consistent with that obtained from U-shaped sources. From a fit to  the
closed-loop source correlations only,  we obtained $Ma=1.12(8)$. For the
other exotic states,  the loop was taken to be in a $L^{PC}=1^{-+}$
representation and the resulting correlations were very noisy compared
to those obtained from U-shaped  sources.

We express our lattice results for the hybrid masses as a ratio to  the
vector meson mass we find on the same lattices, namely $M_V a=0.54(1)$.
Since for the hybrid mesons, we find $Ma \approx 1 $, we must address
possible order $a$ errors which do  not cancel in this mass ratio as $a$
is decreased to the continuum limit. The SW-clover fermionic action has
been used  with the intention of reducing such errors. Indeed a
study~\cite{shan} varying $a$,  shows that dimensionless mass ratios 
are consistent with constant over  a range of lattice spacings including
that used here, unlike the case when the  Wilson fermionic action is
used. It is difficult to estimate the systematic error  from this
extrapolation to the continuum limit of our results without repeating 
the calculation at smaller $a$.  It is  plausible that it is no larger
than the statistical errors we quote.


Another potential source of error is that we use a finite spatial volume
of extent approximately 1.6~fm. Investigations using larger volumes have
shown that this  is large enough for conventional mesons.  For
glueballs, finite size effects are known to be unimportant for  spatial
extent greater than 1 fm.  The static  quark analysis of hybrid mesons,
however, finds that the effective  potential is rather  flat versus
inter-quark separation so that the wave function  will be spatially
extended~\cite{liv,cmper}. This implies that hybrid mesons may be
significantly  more extended than conventional quark model mesons or
glueballs.  Without repeating the hybrid mass calculation  for larger
volumes, we are unable to quantify the magnitude of finite size effects.

For computational reasons, we have used a quark mass which is  heavier
than the strange quark mass  in order to improve the signal to noise
ratio. Ideally one should study hybrid  mesons at a range of quark
masses and interpolate to the experimentally relevant  cases. Here we
use instead a model assumption to make  the small correction to our
quark mass to bring it to the strange quark mass. This is not without
some uncertainty as the model is based on experimental data whereas the
lattice results apply to the quenched approximation - see ref.~\cite{J} 
for a discussion of this problem. In the quenched approximation,  there
will be  magic mixing (i.e. no mixing) and  the $s\bar{s}$ pseudoscalar
will be at 0.68 GeV: thus  $M_{\phi}/M_{\eta(s)}=1.5$.  We find
$M_V/M_P=1.31(1)$ which implies that our quark mass is somewhat heavier
than the strange quark. To correct for this,  the simplest assumption,
which is string inspired,   is that for two types of quark $q_1$
and $ q_2$: $ M(q_1)^2 - M(q_2)^2 $  is the same for 
hybrid, vector and pseudoscalar mesons. This implies that if our lattice
ratio is $M/M_V=x$ then for $s$-quarks $M_s/M_{\phi}=\sqrt{x^2 +0.33
(x^2-1)}$. This correction  is shown in the Table. 

Using the experimental $\phi$ meson mass to set the scale, our result
for the lightest $s \bar{s}$ hybrid meson is 1.99(13) Gev. Taking
account of some of the  possible systematic errors described above, we
round this prediction to 2.0(2) GeV. We can use the  model described
above to estimate that the corresponding non-strange hybrid meson would
be 120 MeV lighter than the $s \bar{s}$ state. 

It is interesting to compare our result with model estimates.   There
have been two classes of model which have been proposed to  describe the
hybrid spectrum. Constituent gluon models, of which the bag 
model~\cite{bag} is the prototype, have a lowest gluonic excitation
which is `magnetic' with $L^{PC}=1^{+-}$. This implies that the
$J^{PC}=1^{-+}$  spin exotic hybrid will be lightest in constituent
models. The spectrum of these `constituent gluons'  has been studied on
a lattice~\cite{gluelump} which confirms that the magnetic gluon is
indeed the lightest such excitation. The other class of model treats the
gluonic excitation as an excited state of a string or  fluxtube, as
motivated by the lattice description~\cite{liv} of hybrid mesons with
heavy quarks. A continuum version of this fluxtube model~\cite{IP}  has
been used to give predictions for hybrid mesons containing light quarks.
These levels are expected to be relatively heavy: about 2.1 GeV for the
$s \bar{s}$ system. This energy value comes basically from the string
model which gives excited  levels with energies $n\pi/R$ higher for
integer $n$. Then $R$, which is the interquark separation, can be
estimated from standard quark models.  This fluxtube model leads to
degenerate  hybrid levels which include  spin-exotic $J^{PC}=1^{-+},\
0^{+-}$, and $2^{+-}$. A reasonable compromise from these two models is 
that the fluxtube model  would be a good approximation to the hybrid
spectrum but that the degeneracy of the spin-exotic levels  would be
broken in the direction predicted by the constituent aproach. This is
exactly what our results indicate.

A study of light-quark  hybrid mesons needs input from the theoretical 
mass values in the quenched approximation to guide experimental searches
and to provide input to models for the  mixing between the $q \bar{q}$
hybrid states and $q \bar{q} q \bar{q}$ or two meson states. Here we
provide the first results of such a study from first principles. We
focus on the spin-exotic states to avoid mixing with $q \bar{q}$ mesons.
For these, the $J^{PC}=1^{-+}$ state is  consistent with being the
lightest. The mass value that we obtain is  2.0(2) GeV for the $s
\bar{s}$ state. It is plausible that this error estimate includes the
systematic errors but these systematic errors can be fully controlled by
extending the lattice study to larger lattices,  smaller  lattice
spacings and several quark masses. Our result for the mass spectrum of
spin-exotic hybrid mesons is  comparable to that suggested by fluxtube
models.

We acknowledge support from EPSRC grant GR/K/41663 to the UKQCD 
Collaboration.  We thank  Stephen Pickles and David Richards for their
help  in setting up the inversion codes for the shifted local
propagators.

\end{document}